\pdfoutput=1
\documentclass[lettersize,journal]{IEEEtran}

\usepackage{xcolor,soul,framed} 

\colorlet{shadecolor}{yellow}
\usepackage{graphicx}
\graphicspath{{pdf/}}
\DeclareGraphicsExtensions{.pdf,.jpeg,.png}

\usepackage[cmex10]{amsmath}
\usepackage{algorithmicx}
\usepackage{algorithm}
\usepackage{array}
\usepackage{mdwmath}
\usepackage{mdwtab}
\usepackage{eqparbox}
\usepackage{url}
\usepackage{amssymb}
\usepackage{algpseudocode}

\usepackage{float}
\usepackage{subfigure}
\usepackage{cite}
\begin{document}
\bstctlcite{IEEEexample:BSTcontrol}
    \title{Mobile Edge Deployment and Resource Management for Maritime Wireless Networks}
  \author{
  Chaoyue Zhang, \textit{Student Member, IEEE}, Bin Lin, \textit{Senior Member, IEEE}, \\ Ziru Chen, \textit{Graduate Student Member, IEEE}, Lin X. Cai, \textit{Senior Member, IEEE}, and Jianli Duan
  \thanks{Manuscript received 23 April 2024; revised 25 July 2024; accepted 17 December 2024. The work of Bin Lin was supported in part by the National Natural Science Foundation of China (No. 62371085, No. 51939001) and in part by the Fundamental Research Funds for the Central Universities (No. 3132023514). The associate editor coordinating the review of this article and approving it for publication was Dr. Phone Lin. \textit{(Corresponding author: Bin Lin.)}}
  \thanks{Chaoyue Zhang and Bin Lin are with the Information Science and Technology College, Dalian Maritime University, Dalian 116026, China (e-mail: zcy\_11335577@163.com; binlin@dlmu.edu.cn).}
  \thanks{Ziru Chen and Lin X. Cai are with the Department of Electrical and Computer Engineering, Illinois Institute of Technology, Chicago, IL 60616, USA (e-mail: zchen71@hawk.iit.edu; lincai@iit.edu).}
  \thanks{Jianli Duan is with the School of Science, Qingdao University of Technology, Qingdao 266520, China (e-mail: duanjianli@qut.edu.cn).}
  }


\maketitle

\begin{abstract}
Mobile Edge Computing (MEC) has been envisioned as one of the key technologies for supplying computation and storage resources in Internet of Vessels (IoV) networks. Due to its flexible deployment, low cost and agile maneuverability, Unmanned Surface Vehicle (USV) has emerged as a promising solution, to provide communication and computation services for maritime users. In this paper, we study mobile edge deployment and resource management for MEC-assisted maritime wireless networks where USVs with diverse computation resources are deployed to provide edge computing services that complement the cloud-based services. To this end, we formulate an optimization problem to minimize the expected response time by jointly optimizing the deployment of mobile USVs and computation offloading decisions. To solve the mixed-integer nonlinear program problem, we propose a Dual-Layer Reinforcement Learning (DLRL) framework to attain a near-optimal solution. Specifically, a Deep Deterministic Policy Gradient (DDPG) algorithm is designed to obtain the best USV deployment in the outer layer learning, and a Q-learning algorithm is designed to determine the best computation offloading decisions in the inner layer learning. Numerical results demonstrate that the proposed solution outperforms some literature algorithms by effectively handling both continuous and discrete variables.
\end{abstract}


%



\section{Introduction}

\IEEEPARstart{I}{nternet} of Vessels (IoV) becomes an indispensable part of maritime wireless communication networks, which enables many smart maritime applications, such as environment monitoring~\cite{r52}, intelligent navigation and voice/video streams, etc. The rapid growth in requirements for maritime applications leads to an increased demand for high bandwidth, low latency, and enhanced computation capabilities. Both academia and the maritime industry conduct research on the deployment of maritime wireless networks to meet the increasing demand from IoV users. Maritime wireless communication networks are typically classified into shore-based communication networks and offshore-based communication networks. Shore-based communication networks can utilize 3G/4G/5G communication technologies to provide broadband data transmissions, typically covering a radius of up to thirty kilometers from the shore. For offshore-based communication networks, communication services can be provisioned through High Frequency (HF) and Very High Frequency (VHF) channels, and maritime satellites. Typical offshore communication systems, such as PACTOR system, Automatic Identification System (AIS) and VHF Data Exchange System (VDES), operate in the HF/VHF band~\cite{r28}. These systems usually offer low data rate services over a large communication coverage, e.g., several hundreds of kilometers. In contrast, maritime satellites can offer high data rate services worldwide from anywhere~\cite{r29}. However, the high cost of vessel-borne equipment and communication expenses pose significant challenges for small vessels. In general, existing maritime wireless networks may not meet the application requirements for communication and computation resources in IoV. 

Mobile Edge Computing (MEC) provides proximate access to users with computation resources through the deployment of edge servers~\cite{r3,r11}. Compared with cloud computing, MEC consumes less transmission time and energy due to a shorter transmission distance~\cite{r12,r61,r62}. In terrestrial networks, MEC has received extensive attention~\cite{r10,r14,r15,r72}. However, research on MEC-assisted maritime wireless networks is lagging behind that in the terrestrial networks. Applying MEC in maritime wireless networks presents new research challenges compared to those in terrestrial networks. First, despite the vast expanse of the ocean surface, finding a stable area for deploying an edge server is a challenging task. Second, maintenance and management of edge servers in maritime environments, such as applying security patches, troubleshooting, and hardware repairs, may incur high operating costs. Third, it is challenging to ensure low-latency transmission from maritime users to the edge due to the hostile communication environment of the ocean. Owing to these complex factors, transmission efficiency is often low in maritime environments. 

In this context, Unmanned Surface Vehicles (USVs) can serve as portable edge servers in close proximity to maritime users, leveraging the advantages of high mobility and ever-reducing cost~\cite{r43,r25,r26}. Given the limited computation resource of an individual USV, how to deploy multiple USVs and design offloading strategies to improve the computation resource utilization, to reduce the computation overload, and to reduce the response time of processing tasks is a critical issue in MEC-assisted maritime wireless networks. It is worth noting that the USV deployment and computation offloading decisions are mutually coupled. Different deployments of USVs may lead to different offloading decisions to achieve the minimum response time, and vice versa. Therefore, USV deployment and computation offloading decisions need to be jointly optimized to minimize the response time. To the best of our knowledge, there have been no prior studies on jointly optimizing multi-USV deployment and computation offloading based on the demands of maritime users in an MEC-assisted maritime wireless network.

In this paper, we formulate a joint optimization problem for USV deployment and computation offloading decisions in MEC-assisted maritime wireless networks. Our objective is to minimize the expected response time, which is comprised of the uplink transmission time, the computation time at the edge server or cloud, and the downlink transmission time. The formulated problem is a Mixed-Integer Non-Linear Program (MINLP) problem with both continuous and discrete variables, i.e., the deployment locations of USVs in a continuous space and the binary offloading decisions in a discrete space. It is worth noting that the deployment locations and the computation offloading decisions are tightly coupled, i.e., the best offloading decisions are dependent on the locations of USVs, and vice versa. In addition, the hybrid variables in both continuous and discrete space pose significant challenges in the optimization framework, as MINLP is well known to be NP-hard. To the best of our knowledge, existing solutions typically involve discretizing continuous actions~\cite{r82,r83}, or employing continuous functions to approximate discrete actions~\cite{r70,r81}, or integrating different tools to handle different types of variables, e.g., to utilize learning techniques for continuous variables and matching theory for discrete variables~\cite{r80,r84}. These solutions may result in reduced accuracy and/or increased computational complexity. To this end, we first reformulate the optimization problem as a Markov Decision Process (MDP), and propose a novel Dual-Layer Reinforcement Learning (DLRL) framework that integrates Deep Deterministic Policy Gradient (DDPG) and Q-learning to handle the complex hybrid action spaces. Specifically, a DDPG algorithm is designed to optimize the locations of USVs, and a Q-learning algorithm is designed to determine the best computation offloading decisions. Finally, extensive simulation results verify that the proposed DLRL algorithm can effectively reduce the expected response time, in comparison with the benchmark algorithms in the literature. The main contributions of this work can be summarized as follows.

\begin{itemize}
\item We propose a framework for mobile edge deployment and resource management in MEC-assisted maritime wireless networks, where computation tasks of each vessel can be offloaded to an edge server or cloud center. A joint USV deployment and computation offloading problem is formulated to minimize the expected response time, subject to the constraint of edge computation resources.
\end{itemize}

\begin{itemize}
\item 
To solve the formulated optimization problem, we reformulate it as an MDP problem, and propose a DLRL framework that integrates DDPG and Q-learning to handle the discrete and continuous actions to facilitate collaborative training. 
\end{itemize}

\begin{itemize}
\item In the DLRL algorithm, the outer layer DDPG is designed to optimize the locations of USVs in continuous space, based on the locations and computation tasks of vessels from historical experiences; and the inner layer Q-learning is designed to determine the best computation offloading decisions in discrete space based on the locations of USVs. Further analysis is conducted on the complexity of the proposed learning algorithm.
\end{itemize}

The remainder of this paper is organized as follows. The related works are presented in Section~\ref{II}. An MEC-assisted maritime wireless network model and problem formulation are introduced in Section~\ref{III}. A DLRL algorithm is proposed to solve the optimal USV deployment and computation offloading problem in Section~\ref{IV}. Numerical results are presented to evaluate the performance of the proposed DLRL algorithm in Section~\ref{V}, followed by concluding remarks in Section~\ref{VI}.


\section{Related Works} \label{II}
Computation offloading in MEC networks has been extensively studied in the literature~\cite{r72,r74,r73}. Generally, the computation tasks can be processed at mobile users, or be offloaded to the edge server or the the cloud server. In~\cite{r72}, a DDQN algorithm is applied to jointly optimize the computation offloading decisions, and spectrum and computation resource allocation in an MEC network with a single edge server, aiming to minimize the energy consumption of the system. Computation offloading in an MEC network with multiple edge servers is studied in~\cite{r74,r73}. In~\cite{r74}, convex optimization methods are used to optimize the computation offloading, the computation resource allocation, and transmission power control in an MEC network, with the objective of minimizing the overall energy consumption. In~\cite{r73}, an end-to-end deep reinforcement learning algorithm is proposed to optimize the offloading decisions and computational frequency allocation in an MEC network with multiple users and multiple edge servers, aiming to maximize the number of tasks completed on time while minimizing the energy consumption.
 
The deployment of edge servers can be categorized into two types, i.e., 1) fixed deployment, where edge servers are strategically placed within the network infrastructure, with the location unchanged after deployment to ensure stable and reliable edge services; and 2) on-demand deployment, where edge servers are deployed based on network dynamics, and the configuration of the edge servers is adjustable in response to the changing network conditions and demands.

Some studies investigate deploying edge servers among a set of fixed candidate locations. In~\cite{r65}, leveraging the advantages of the bipartite graph and heuristic algorithm, a two-step optimization method is proposed to address the edge server deployment and task scheduling, aiming to maximize the profits of the service provider. In~\cite{r66}, a deployment algorithm based on k-means clustering and particle swarm optimization is proposed to solve the cloudlet deployment problem, with the objective of minimizing the average service delay. In these works, a single objective optimization is formulated. There are some works that formulate the edge server deployment problem as a multi-objective optimization problem\cite{r10,r8,r9}. In~\cite{r8}, the objective is to balance the workload and minimize the access delay. Genetic algorithm and local search algorithm are designed to find the best solution. The work in~\cite{r9} studies how to strike a trade-off among the coverage rate, the waiting time, and the workload balance. The Niched Pareto Genetic Algorithm II (NPGA-II) is adopted to find the feasible solutions for edge server deployment. In these aforementioned works, heuristic algorithms are proposed to find the deployment solutions, and the deployment is not adaptive to the dynamic demands of the network.

Recent advancement on Unmanned Aerial Vehicles (UAVs) offers a viable and flexible solution for MEC where UAVs equipped with edge servers can provide on-demand services to users in the proximity and the locations of UAVs can dynamically adapt to the user demands. The trajectories of UAVs are usually modeled as continuous variables, and efficient methods and algorithms based on machine learning and convex optimization are proposed to dynamically optimize decision-making processes. The deployment of a single UAV equipped with an edge server is studied in~\cite{r67,r68}. In~\cite{r67}, an alternative algorithm based on convex optimization and Lyapunov-based approach is proposed to optimize the resource allocation and UAV trajectory in a UAV-assisted MEC system, with the objective of minimizing the average energy consumption. In~\cite{r68}, an alternative algorithm based on block coordinate descent and successive convex approximation techniques is proposed to solve the computation offloading decisions, bits allocation and UAV trajectory optimization problem in a UAV-enhanced MEC network, with the objective of minimizing the total energy consumption. A few research works study multi-UAV deployment and computation offloading in the multi-UAV-enabled MEC network~\cite{r70,r87,r86,r57}. Generally, joint optimization of multiple UAV trajectories and offloading decisions in a wireless network is a complex optimization problem which is usually modeled as an MINLP problem~\cite{r69}. Deep reinforcement learning offers an efficient solution to tackle the challenging problem. In~\cite{r70}, a Q-learning algorithm and transport theory are proposed to optimize the multi-UAV deployment and user association, with the objective of minimizing the transmit power consumption. The work in~\cite{r87} develops a multi-agent DDPG algorithm to jointly optimize the vehicle association, resource allocation and multi-UAV trajectory in an MEC- and UAV-assisted vehicular network. The algorithms used in~\cite{r70,r87} utilize the approximation method to handle the complex optimization problem, resulting in performance loss. The work in~\cite{r86} applies a many-to-one matching to optimize the subchannel allocation, and leverages a DDPG algorithm to train the multi-UAV trajectory in a multi-UAV collaborative caching network, aiming to minimize the total content retrieving delay. In~\cite{r57}, an algorithm based on the transport theory and particle swarm intelligence is proposed to find the optimal user association and multi-UAV deployment, with the objective of minimizing the average task delay. The algorithms used in~\cite{r86,r57} utilize iteration optimization to separately solve the continuous and discrete variables to attain a local optimal solution with high computational complexity. In addition, the existing works mainly consider a small scale MEC network with no more than four UAVs. The complexity of these algorithms usually increases exponentially with the number of UAVs, and thus they are not suitable for a medium or a large scale deployment of MEC with a large number of mobile edge servers.

Similar to UAVs, USVs equipped with edge servers can provide mobile edge services to users in a maritime wireless network. Joint resource management and USV deployment are studied in~\cite{r38,r59}, and successive convex approximation, semidefinite programming relaxation, interior-point and block coordinate descent methods are proposed to solve the joint optimization problem. Numerical results show that a USV deployed in the maritime wireless network can greatly improve the network throughput. Multi-USV deployment is studied in~\cite{r75}, where bisection search is used to find the optimal placement of USVs along with the data uploading duration and signal powers to achieve the minimum energy consumption. In this work, the number of USVs is no more than six. 

The aforementioned literature works focus on either computation task offloading or edge server deployment in an MEC network. Basically, offloading decisions should be adaptive to the dynamic demands of users, which is not generally coupled with network deployment in traditional MEC with fixed deployment of edge servers. Thanks to the mobility of USVs, it is possible and favorable to deploy mobile edge servers according to the dynamic network demands. Therefore, a joint optimization of the edge server deployment and computation offloading beckons for further research. In addition, while most existing works only consider a limited number of USVs/UAVs, a scalable solution is required for a maritime wireless network because a large number of USVs are required to cover a wide area in marine. To the best of our knowledge, our work is the first to jointly optimize multi-USV deployment and computation offloading based on the demands of maritime users in an MEC-assisted maritime wireless network.

\section{System Model and Problem Formulation} \label{III}
\begin{table}[t] 
\renewcommand{\arraystretch}{1.2}
\caption{Main Notations and Descriptions}\label{tab1}
\centering
\begin{tabular}{|p{1.2cm}|p{6.8cm}|}
\hline
\textbf{Notation} & \textbf{Description}\\
\hline
$\mathcal{U}$ & A set of USVs\\
\hline
$\mathcal{M}$ & A set of vessels\\
\hline
$\mathcal{N}$ & A set of USV clusters\\
\hline
$\mathcal{Z}_n$ & A set of USVs in USV cluster $n$\\
\hline
$\alpha_{m,n}$ & Offloading decision variable. 1 means vessel $m$ offloads task to USV cluster $n\in \mathcal{N}$ or the cloud center $n=\{0\}$; and 0 vice versa.\\
 \hline
$\lambda_{m}$ & The task arrival rate of vessel $m$ \\
\hline
$I_{m}$ & The number of computation tasks of vessel $m$ \\
\hline
$c_{m}$ & The required CPU cycles for processing the computation task of vessel $m$ \\
\hline
$b_{m}^{\rm in}$ & The average input data volume of the computation task of vessel $m$ \\
\hline
$b_{m}^{\rm out}$ & The average data volume of the computation task computation result of vessel $m$ \\
\hline
$\mathbf{L}_i$ & The location of $i$, where $i \in \mathcal{M} \cup \mathcal{N}$ \\
\hline
$P_i$ & The transmit power of $i$, where $i \in \mathcal{M} \cup \mathcal{N}$ \\
\hline
$H_i$ & The antenna height of $i$, where $i \in \mathcal{M} \cup \mathcal{N}$ \\
\hline
$B$ & Channel bandwidth \\
\hline
$h_{i,j}$ & The channel gain between $i$ and $j$, where $i,j \in \mathcal{M} \cup \mathcal{N}$ \\
\hline
$R_{i,j}$ & The transmission rate from $i$ to $j$, where $i,j \in \mathcal{M} \cup \mathcal{N}$ \\
\hline
$C_n$ & The computation resource of $n$, where $n \in \mathcal{N} \cup \{0\}$ \\
\hline
\end{tabular}
\end{table}

\subsection{Network Model}
\begin{figure*}[t]
  \begin{center}
  \includegraphics[width=7in]{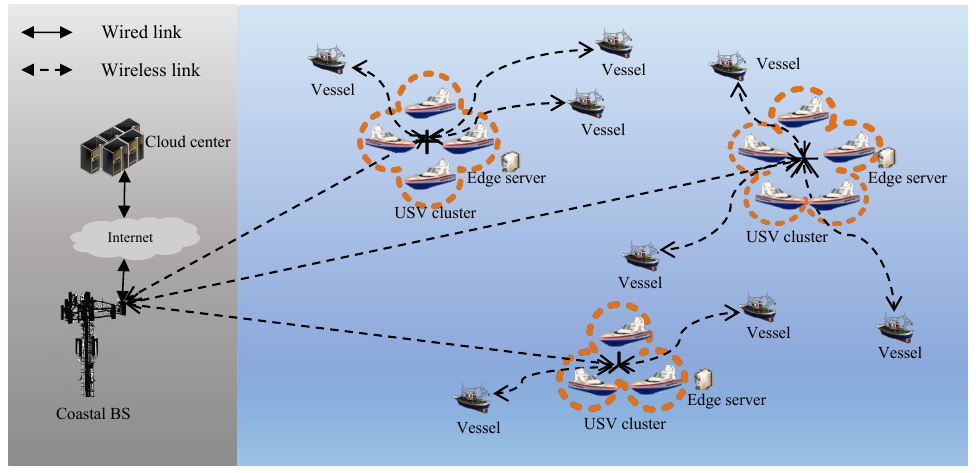}\\
  \caption{An MEC-assisted maritime wireless network with multiple USVs.}\label{fig1}
  \end{center} 
\end{figure*} \label{fig:network}
In this paper, we consider an MEC-assisted maritime wireless network consisting of a cloud center, a coastal Base Station (BS), a group of USVs denoted as $\mathcal{U}=\{1,...,U\}$ and a group of vessels denoted as $\mathcal{M}=\{1,...,M\}$, as illustrated in Fig.~\ref{fig1}. Each vessel is associated with a USV via a wireless link, and USVs can communicate with a coastal BS, which is connected to the cloud center via a wired link. The BS serves as the network controller that collects the network information for mobile edge deployment and resource management. Each USV can be equipped with computation and storage units to provision edge computing services for vessels in the proximity. Due to the limited computation resources in each individual USV, multiple USVs can be deployed in one area to form a cluster of USVs so that the vessel associated with the USVs in the cluster can share the aggregated computing resources. Without loss of generality, USVs are grouped into $N$ clusters, and each USV cluster may contain one\footnote{This is the special case that each USV will be individually deployed without clustering.} or multiple USVs. Denote the set of USVs in USV cluster $n \in \mathcal{N}=\{1,...,N\}$ as $\mathcal{Z}_n=\{1,...,Z_n\}, \forall z_n \in\mathcal{Z}_n$, where $\sum_{n=1}^N Z_n=U (U \geq N)$ USVs. In a 2D Cartesian coordinate system, the location of $i$ is denoted as $\mathbf{L}_i=[x_i,y_i]$, where $i$ can be a vessel $i \in \mathcal{M}$, or a USV cluster $i \in \mathcal{N}$. Assume that the task arrival rate of vessel $m$ follows Poisson distribution with mean $\lambda_{m}$. Let $I_{m}$ denote the number of computation tasks of vessel $m$ in a time slot. The main notations are listed in Table~\ref{tab1}.

Considering the impacts of sea surface reflection and antenna height, a two-ray signal transmission model is adopted for maritime channels~\cite{r29}. The transmission model assumes that the maritime channel mainly consists of a direct path and a reflection path, the channel gain between vessel $m$ and the USV cluster $n$ can be expressed as
\begin{equation}
\begin{aligned}
    h_{m,n} &= {\left({\frac{l}{4\pi d_{m,n}}}\right)^2 \left[2{\rm sin} \left(\frac{2\pi {H_m}{H_n}}{ld_{m,n}} \right) \right]^2},\\ & \quad\quad \forall m \in \mathcal M, \forall n \in \mathcal N,
\end{aligned}
\end{equation}
where $l$ is the carrier wavelength, $H_m$ is the antenna height of vessel $m$, $H_n$ is antenna height of $n$-th USV cluster header, $d_{m,n}=\sqrt{(x_m-x_n)^2+(y_m-y_n)^2}$ is the communication link distance. The uplink transmission rate $R_{m,n}$ from vessel $m$ to USV cluster, and the downlink transmission rate $R_{n,m}$ from USV cluster $n$ to vessel $m$ are
\begin{equation}
	\begin{aligned}
		R_{m,n} = B \log_2 \left(1+ \frac{P_m h_{m,n}}{\sigma^2} \right), \forall m \in \mathcal M, \forall n \in \mathcal N,
	\end{aligned}
\end{equation}
and
\begin{equation}
	\begin{aligned}
		R_{n,m} = B \log_2 \left(1+ \frac{P_n h_{m,n}}{\sigma^2} \right), \forall m \in \mathcal M, \forall n \in \mathcal N,
	\end{aligned}
\end{equation}
respectively, where $\sigma^2$ denotes the white Gaussian noise power, $B$ denotes the channel bandwidth, $P_m$ denotes the transmit power of vessel $m$, $P_n$ denotes the transmit power of $n$-th USV cluster header.

Vessels can offload their tasks to USV clusters or the cloud center. We introduce a binary variable $\alpha_{m,n}\in \{0,1\}, \forall m \in \mathcal M, \forall n \in \mathcal N \cup \{0\}$ to represent computation offloading decision of vessel $m$. Specifically, $\alpha_{m,n}=1$ indicates that vessel $m$ offloads the task to a USV cluster $n\in \mathcal{N}$ or the cloud center $n=\{0\}$, and $\alpha_{m,n}=0$ otherwise. Each vessel can offload its computation tasks to either one USV cluster for edge computing or the cloud center for cloud computing, satisfying
\begin{equation}
    \sum_{n\in \mathcal{N}\cup \{0\}} \alpha_{m,n} = 1, \forall m \in \mathcal{M}.
\end{equation}

\emph{Case 1: Edge Computing.} In the case that the computation tasks are offloaded to the edge server, the response time of the vessel is the sum of the uplink transmission time from vessel $m$ to a USV cluster $n$, denoted as $T_{m,n}^{\rm ul}$, the computation time at the USV cluster $n$, denoted as $T_{m,n}^{\rm comp}$, and the downlink transmission time from a USV cluster $n$ to vessel $m$, denoted as $T_{n,m}^{\rm dl}$. Thus, we have 
\begin{equation}
    T_{m,n}=T_{m,n}^{\rm ul} + T_{m,n}^{\rm comp} + T_{n,m}^{\rm dl}, \forall m \in \mathcal{M}, \forall n \in \mathcal{N}.
\end{equation}

The uplink transmission time $T_{m,n}^{\rm ul}$ is given by
\begin{equation}
    T_{m,n}^{\rm ul}=\alpha_{m,n}\frac{I_{m}b_{m}^{\rm in}}{R_{m,n}}, \forall m \in \mathcal{M}, \forall n \in \mathcal{N},
\end{equation}
where $b_{m}^{\rm in}$ denotes the average input data volume of the computation task of vessel $m$.

The computation resource of USV cluster $n$ is $C_n=\sum_{u\in\mathcal{Z}_n}C_{u}$, where $C_{u}$ is the computation resource equipped in USV $u$. The computation time at USV cluster $n$ to process the offloaded tasks of vessel $m$ is
\begin{equation}
T_{m,n}^{\rm comp}=\alpha_{m,n}\frac{I_{m}c_{m}}{C_n}, \forall m \in \mathcal{M}, \forall n \in \mathcal{N},
\end{equation}
where $c_{m}$ denotes the required CPU cycles for computing the computation task of vessel $m$.

The downlink transmission time of computation results to be sent from USV cluster $n$ to vessel $m$ is given by
\begin{equation}
    T_{n,m}^{\rm dl}=\alpha_{m,n}\frac{I_{m}b_{m}^{\rm out}}{R_{n,m}}, \forall m \in \mathcal{M}, \forall n \in \mathcal{N},
\end{equation}
where $b_{m}^{\rm out}$ denotes average data volume of the task computation result of vessel $m$.

Note that the total edge computation resource to serve vessels cannot exceed the maximum computation resource of the USV cluster $n$, 
\begin{equation}
    \sum_{m\in\mathcal{M}}\alpha_{m,n}\lambda_{m}c_{m}\leq C_n, \forall n \in \mathcal{N}.
\end{equation}

\emph{Case 2: Cloud Computing.} 
In the case that the computation tasks are offloaded to the cloud, the response time is the sum of the uplink transmission time from vessel $m$ to the cloud center, denoted as $T_{m,0}^{\rm ul}$, the computation time at the cloud center, denoted as $T_{m,0}^{\rm comp}$, and the downlink transmission time from the cloud center to vessel $m$, denoted as $T_{0,m}^{\rm dl}$. Thus, we have

\begin{equation}
    T_{m,0}=T_{m,0}^{\rm ul} + T_{m,0}^{\rm comp} + T_{0,m}^{\rm dl}, \forall m \in \mathcal{M}.
\end{equation}

The uplink transmission time $T_{m,0}^{\rm ul}$ is given by
\begin{equation}
\begin{aligned}
    T_{m,0}^{\rm ul}=\alpha_{m,0}\left(\frac{I_{m}b_{m}^{\rm in}}{R_{m,n}} + \frac{I_{m}b_{m}^{\rm in}}{R_{n,0}}+ \tau_0 \right),\forall m \in\mathcal{M}, \forall n \in \mathcal{N},
\end{aligned}
\end{equation}
where the first term in the braces represents the task transmission time from vessel $m$ to the nearest USV cluster $n$, the second term is the transmission time from the USV cluster $n$ to the coastal BS, and the third term $\tau_0$ denotes the transmission time from the coastal BS to the cloud center.

Generally, the computation resource of the cloud center is much greater than that in the edge server, and we have $C_0 >> C_n$. The computation time at the cloud center to process the offloaded tasks of vessel $m$ is
\begin{equation}
    T_{m,0}^{\rm comp}=\alpha_{m,0}\frac{I_{m}c_{m}}{C_0}, \forall m \in\mathcal{M}.
\end{equation}

The computation results are then transmitted from the cloud center to vessel $m$ through the coastal BS, and the transmission time is thus given by
\begin{equation}
\begin{aligned}
    T_{0,m}^{\rm dl}=\alpha_{m,0}\left(\tau_0 + \frac{I_{m}b_{m}^{\rm out}}{R_{\rm 0,n}} + \frac{I_{m}b_{m}^{\rm out}}{R_{\rm n,m}}\right), \forall m \in\mathcal{M}, \forall n \in \mathcal{N},
\end{aligned}
\end{equation}
where the first term in the braces, $\tau_0$, denotes the transmission time of computation results from the cloud center to the coastal BS, the second term represents the transmission time from the coastal BS to the USV cluster $n$, and the third term represents the transmission time from the USV cluster $n$ to vessel $m$.

\subsection{Problem Formulation}
The objective is to minimize the expected response time by jointly optimizing the deployment of USV clusters $\mathbf{L}^{\rm usv}=\{\mathbf{L}_n, \forall n \in \mathcal{N}\}$ and the computation offloading decisions for vessels $\boldsymbol{\alpha}=\{\alpha_{m,n}, \forall m \in \mathcal{M}, \forall n \in \mathcal{N} \cup \{0\}\}$, while satisfying the edge computing resource constraint. Mathematically, we formulate the Expected Response Time Minimization (ERTM) problem as follows,
\begin{subequations}
\begin{align*}
    &{\rm (\textbf{ERTM}):}\\
	\min_{\mathbf{L}^{\rm usv},\boldsymbol{\alpha}} \
    &\quad T=\frac{1}{M}\sum_{m\in\mathcal{M}}(T_{m,n}+T_{m,0}) \tag{14} \\
	\text{s.t.}
	&\quad \alpha_{m,n} \in \{0,1\}, \ \forall m \in \mathcal M, \forall n \in\mathcal{N}\cup\{0\}, \tag{14a} \label{eq:15a}\\
     &\quad \sum_{n\in\mathcal{N}\cup \{0\}}\alpha_{m,n} = 1, \forall m\in\mathcal{M}, \tag{14b} \label{eq:15b}\\
    &\quad \sum_{m\in\mathcal{M}}\alpha_{m,n}\lambda_{m}c_{m} \leq C_n, \forall n \in\mathcal{N}, \tag{14c}  \label{eq:15c} \\
    &\quad 0\leq x_n \leq x_{\rm max}, 0 \leq y_n \leq y_{\rm max}, \forall n \in\mathcal{N}. \tag{14d}  \label{eq:15d}\\
\end{align*}
\end{subequations}
Constraints (\ref{eq:15a}) and (\ref{eq:15b}) represent that each vessel can offload its tasks to either one USV cluster or the cloud center. Constraint (\ref{eq:15c}) means that the total computation resource cannot exceed the maximum computation resource of the USV cluster. Constraint (\ref{eq:15d}) is the deployment area constraint. 

\section{Dual-Layer Reinforcement Learning Algorithm Design} \label{IV}
In this section, we first reformulate the ERTM problem as an MDP, and then propose a dual-layer reinforcement learning framework to handle the hybrid action spaces and to attain the optimal solution.

\subsection{MDP Formulation} \label{IV-A}
In this subsection, we reformulate the ERTM problem as an MDP. For the DLRL algorithm, the coastal BS is modeled as an agent, and the MEC-assisted maritime wireless network is the environment. The state space, the action space, and the reward function of the MDP are defined as follows.

\textit{1) Action Space:} The action space in the $t$-th round is $\mathbf{a}_t = \{\mathbf{v}_t, \boldsymbol{\vartheta}_t, \mathbf{D}_t\}$, where $\mathbf{v}_t = \{v_{m,t} | m \in \mathcal{M}\}$ denotes the computation offloading decisions of vessels, $\boldsymbol{\vartheta}_t = \{\vartheta_{n,t} | \vartheta_{n,t} \in [0, 2\pi), \forall n \in \mathcal{N}\}$ denotes the search angle for USV clusters, and $\mathbf{D}_t=\{D_{n,t} | D_{n,t} \in[0,D_{\max}], \forall n \in \mathcal{N}\}$ denotes the search distance of USV clusters within a maximal search distance $D_{\max}$.

\textit{2) State Space:} The state space in the $t$-th round is $\mathbf{s}_t = \{\mathbf{L}^{\rm usv}_t, \mathbf{h}_{t}\}$, where $\mathbf{L}^{\rm usv}_t = \{\mathbf{L}_{n,t}|n\in\mathcal{N}\}$ denotes the locations of USV clusters, and $\mathbf{h}_{t} = \{h_{m,n,t}|m\in\mathcal{M}, n\in\mathcal{N}\}$ denotes the channel gain between vessels and USV clusters.

\textit{3) Reward Function:} According to the problem formulation in Section~\ref{III}, the objective of the ERTM problem is to minimize the expected response time by optimizing the deployment of each USV cluster and computation offloading decisions. Therefore, the reward function of the DLRL algorithm should be inversely proportional to the expected response time, which is defined as
\begin{equation}
    r_t = \frac{1}{T(\mathbf{s}_{t},\mathbf{a}_t)},
\end{equation}
where $T(\mathbf{s}_{t},\mathbf{a}_t)$ is the expected response time of all vessels at the state $\mathbf{s}_{t}$ and taking the action $\mathbf{a}_t$.

\subsection{Dual-Layer Reinforcement Learning}
\begin{figure*}[t]
  \centering
  \includegraphics[width=6.6in]{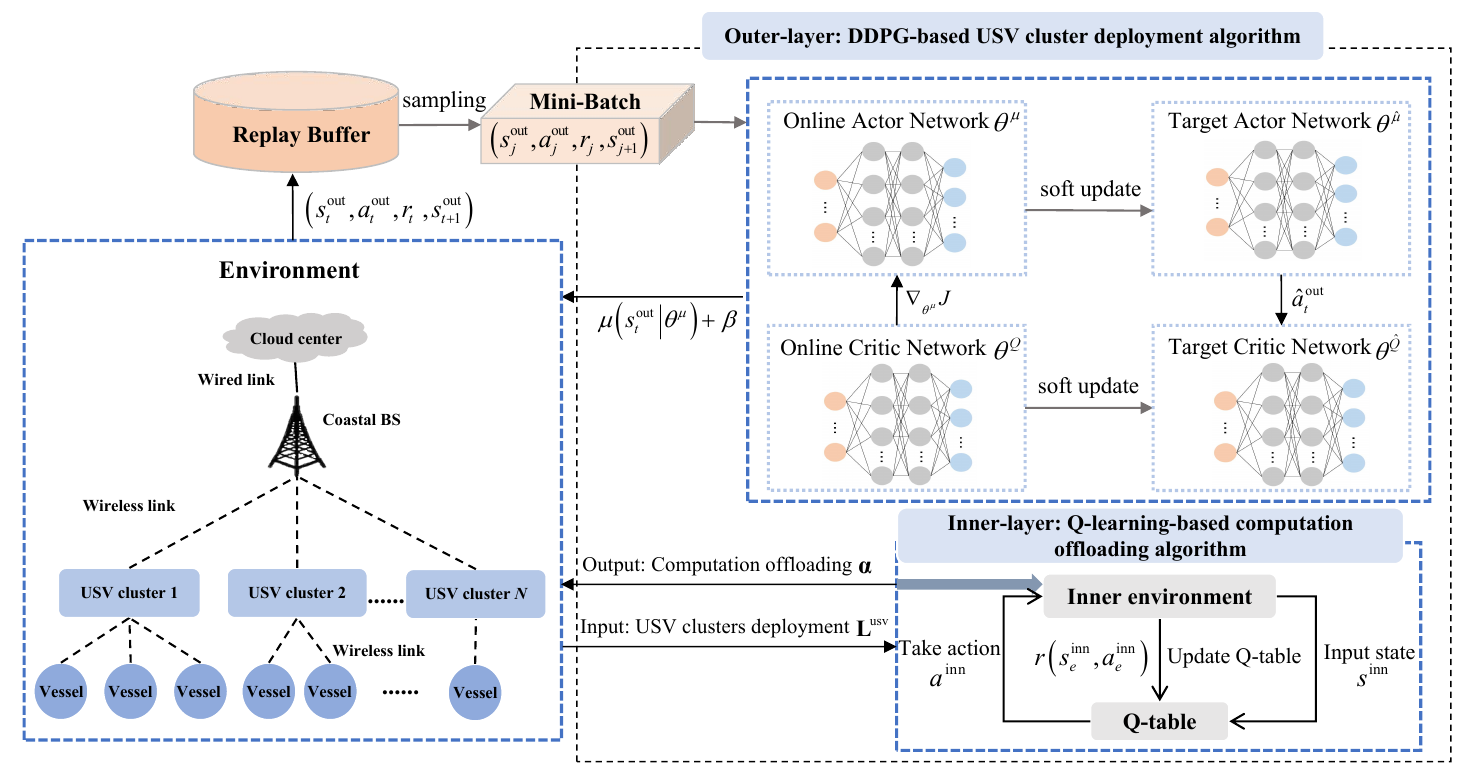}\\
  \caption{Framework of the DLRL algorithm.}\label{fig2}
\end{figure*}

We propose a dual-layer learning framework to solve the formulated problem. As shown in Fig.~\ref{fig2}, the outer layer DDPG is employed for the USV deployment in continuous space and the inner layer Q-learning is employed for the computation offloading decisions in discrete space. Specifically, the inner layer learning is executed at the beginning of each epoch, which can obtain the best computation offloading results and interact with the environment. The results of the inner layer learning is used to further train the outer layer Deep Neural Networks (DNNs). In the outer layer, we utilize a DDPG to train the USV deployment design, where two main DNNs, i.e., the actor network and the critic network, are trained with the same structure but different parameters. In the DLRL, a mini-batch is used to sample the historical experience stored in the replay buffer for updating network parameters. Finally, the USV deployment actions and computation offloading selection actions are coordinated to achieve convergence. The detailed designs of the outer layer and inner layer learning algorithms are presented as follows.

\subsubsection{Outer layer framework of DLRL algorithm} \label{IV-C}
A deep reinforcement learning algorithm based on DDPG is proposed to optimize the deployment locations of USVs in the continuous action space. Based on the MDP, DDPG can guide the agent to efficiently approach the near-optimal solution with continuous variables by exploring and interacting with the environment. The agent observes the current environment state, selects an action, receives an immediate reward based on the current state and action, and then observes the next state. Specifically, for the outer layer, the action space in the $t$-th round is $\mathbf{a}_t^{\rm out}=\{\boldsymbol{\vartheta}_{t}, \mathbf{D}_{t}\}$, and the state space in the $t$-th round is $\mathbf{s}_t^{\rm out}=\{\mathbf{L}_{t}^{\rm usv}, \mathbf{h}_{t}\}$.

The DDPG algorithm consists of two main networks, i.e., the actor network and the critic network. Both the actor and critic networks feature two subnets, i.e., the online network and the target network with the same structure. The critic network is trained to approximate the Q-values based on the state and action. The actor network is trained to generate a deterministic policy and returns the corresponding action based on the current state. Besides, experience replay buffer and target networks are employed to enhance the stability and convergence of the learning process in the online networks.

The goal of DDPG is to find optimal policies that maximize the cumulative reward value. A policy $\mu$ can be considered as a function that maps states to actions. Thus, we can obtain the action-value function as follows,
\begin{equation}
    Q(\mathbf{s}_{t}^{\rm out},\mathbf{a}_{t}^{\rm out}|\theta^Q)=\mathbb{E}[r_t+\zeta^{\rm out} Q(\mathbf{s}_{t+1}^{\rm out},\mathbf{a}_{t+1}^{\rm out}|\theta^Q)],
\end{equation}
where $\theta^Q$ is the weight of the online critic network, $\zeta^{\rm out} \in [0,1]$ is the discount factor that represents the uncertainly of future revenue, $\mathbb{E}[\cdot]$ is the expectation.

The training process can be described as follows. Initially, the actor network adds the noise $\beta$ and generates the action $\mathbf{a}_t^{\rm out}=\mu(\mathbf{s}_t^{\rm out}|\theta^{\mu})+\beta$, where $\theta^{\mu}$ is the weight of the online actor network. Upon interaction with the environment using action  $\mathbf{a}_t$, the agent receives a reward $r_t$ and transitions to the next state $\mathbf{s}_{t+1}^{\rm out}$. This experience $(\mathbf{s}_t^{\rm out},\mathbf{a}_t^{\rm out},r_t,\mathbf{s}_{t+1}^{\rm out})$ is stored in the experience replay buffer $\mathbf{O}$. After that, a set of historical experiences from the replay buffer are randomly sampled into a mini-batch buffer, such as $(\mathbf{s}_j^{\rm out},\mathbf{a}_j^{\rm out},r_j,\mathbf{s}_{j+1}^{\rm out})$,  to update the network weights.

The critic network updates its online network by minimizing loss of the critic network, which is approximated by
\begin{equation} \label{eq20}
    L(\theta^Q) = \mathbb{E}[(Q(\mathbf{s}_j^{\rm out},\mathbf{a}_j^{\rm out}|\theta^Q)-Y_j)^2], 
\end{equation}
where $Y_j=r_j+\zeta Q(\mathbf{s}_{j+1}^{\rm out},\hat{\mu}(\mathbf{s}_{j+1}^{\rm out}|\theta^{\hat{\mu}})|\theta^{\hat{Q}})$ is the Q-value produced by the target critic network, $\theta^{\hat{\mu}}$ and $\theta^{\hat{Q}}$ are weights of the target actor network and the target critic network, respectively. The symbol $\hat{\mu}(\cdot|\theta^{\hat{\mu}})$ means the action policy obtained by the target actor network. Correspondingly, the weight $\theta^Q$ of the online critic network is updated as
\begin{equation} \label{eq21}
\begin{aligned}
    &\bigtriangledown_{\theta^Q}L(\theta^Q) \\
    &=\mathbb{E}[(Q(\mathbf{s}_j^{\rm out},\mathbf{a}_j^{\rm out}|\theta^Q)-Y_j)\times\bigtriangledown_{\theta^Q}Q(\mathbf{s}_j^{\rm out},\mathbf{a}_j^{\rm out}|\theta^Q)],
\end{aligned}
\end{equation}
\begin{equation} \label{eq22}
    \theta^Q = \theta^Q + \gamma_Q \bigtriangledown_{\theta^Q}L(\theta^Q),
\end{equation}
where $\bigtriangledown_{\theta^Q}\cdot$ denotes the gradient vector with the weight ${\theta^Q}$, $\gamma_Q$ is the learning rate of the online critic network.

With the Adam optimizer, the gradient of the weight ${\theta^\mu}$ is derived as $\bigtriangledown_{\theta^\mu}\mu(\mathbf{s}_j^{\rm out}|\theta^\mu)$. Based on the sampling mini-batch, the policy gradient of the online actor network with the weight $\theta^\mu$ is updated as
\begin{equation} \label{eq18}
    \bigtriangledown_{\theta^\mu}J(\theta^\mu)=\mathbb{E}[\bigtriangledown_aQ(\mathbf{s}_j^{\rm out},a|\theta^Q)|_{a=\mu(\mathbf{s}_j^{\rm out}|\theta^\mu)}\bigtriangledown_{\theta^\mu}\mu(\mathbf{s}_j^{\rm out}|\theta^\mu)].
\end{equation}
Correspondingly, the weight $\theta^\mu$ of the online actor-network is expressed as
\begin{equation} \label{eq19}
	\theta^\mu=\theta^\mu+\gamma_\mu\bigtriangledown_{\theta^\mu}J(\theta^\mu), 
\end{equation}
where $\gamma_\mu$ is the learning rate of the online actor network.

With the soft update strategy, the weights of the target actor network and the target critic network are updated as
\begin{equation} \label{eq23}
	\theta^{\hat{\mu}}\leftarrow\delta\theta^\mu+(1-\delta)\theta^{\hat{\mu}},
\end{equation}
and
\begin{equation} \label{eq24}
	\theta^{\hat{Q}}\leftarrow\delta\theta^Q+(1-\delta)\theta^{\hat{Q}},
\end{equation}
respectively. Here, $\delta \in [0,1]$ is the soft update coefficient.

\subsubsection{Inner layer framework of DLRL algorithm} \label{IV-D}
Q-learning is an effective model-free reinforcement learning algorithm based on value iteration, which can find the optimal action in a finite MDP~\cite{r79}. In the inner layer, the action space is $\mathbf{a}^{\rm inn}_e = \{\mathbf{v}_e\}$, and the state space is $\mathbf{s}^{\rm inn}_e = \{\mathbf{h}_e\}$. The agent computes the value of each state-action pair, and updates the Q-value in a Q-table after each interaction. 

The $\epsilon$-greedy policy is employed to strike a balance between exploitation and exploration. Specifically, the agent selects the optimal action based on the current Q-table with a probability of $1-\epsilon$ for exploitation. With a probability of $\epsilon$, the agent selects a random action for exploration, which enables the agent to try out new actions and potentially discover better ones that may not have been selected through exploitation. Thus, the action with $\epsilon$-greedy policy in the $e$-th epoch is given by
\begin{equation}
    \mathbf{a}^{\rm inn}_{e} =
 \left\{\begin{aligned}
        &{\rm argmax}_{\mathbf{a}^{\rm inn}_{e} \in \mathbf{a}^{\rm inn}}Q(\mathbf{s}^{\rm inn}_{e},\mathbf{a}^{\rm inn}_{e}), {\rm with\ probability}\ \epsilon, \\
        &{\rm a\ random\ action}, {\rm with\ probability}\ 1-\epsilon.
       \end{aligned}
\ \right.
\end{equation}

Q-value is a state-action value, which describes the benefit of an agent performing a particular action in a state. Therefore, the ultimate goal of Q-learning is to appropriately update the Q-table such that the agent obtains the maximum reward. The Q-value in the Q-table can be calculated by
\begin{equation} \label{eq25}
\begin{aligned}
    Q(\mathbf{s}^{\rm inn}_{e},\mathbf{a}^{\rm inn}_{e}) \leftarrow & Q(\mathbf{s}^{\rm inn}_{e},\mathbf{a}^{\rm inn}_{e}) + \gamma^{\rm inn} [r(\mathbf{s}^{\rm inn}_{e},\mathbf{a}^{\rm inn}_{e}) + \\
    & \zeta^{\rm inn} \max_{\mathbf{a}^{inn}_{e+1}}Q(\mathbf{s}^{\rm inn}_{e+1},\mathbf{a}^{\rm inn}_{e+1}) - Q(\mathbf{s}^{\rm inn}_{e},\mathbf{a}^{\rm inn}_{e})],
\end{aligned}
\end{equation}
where $Q(\mathbf{s}^{\rm inn}_{e},\mathbf{a}^{\rm inn}_{e})$ is the state-action value, $\zeta^{\rm inn} \in \{0,1\}$ is the discount factor representing the weight of future rewards, and $\gamma^{\rm inn}$ is the learning rate. 

The proposed DLRL algorithm, including the inner layer Q-learnig and outer layer DDPG is presented in Algorithm 1.
\begin{algorithm}[t]
	\caption{DLRL Algorithm}  
	\begin{algorithmic}[1] 
		\Require Episode length $E^{\rm out}$ and $E^{\rm inn}$, training round $T$, experience replay buffer $\mathbf{O}$, mini-batch size $o$, learning rate for actor network $\gamma_\mu$, learning rate for critic network $\gamma_Q$, discount factor $\zeta^{\rm out}$, learning rate of Q-learning $\gamma^{\rm inn}$, soft update coefficient $\delta$;
		\Ensure Locations of USV clusters $\mathbf{L}^{\rm usv}$, computation offloading decisions $\boldsymbol{\alpha}$;
		\State Initialize online actor network and online critic network with the weights $\theta^\mu$ and $\theta^Q$;
		\State Initialize target actor network and target critic network with the weights $\theta^{\hat{\mu}}$ and $\theta^{\hat{Q}}$;
        \State Initialize the locations of USV clusters $\mathbf{L}^{\rm usv}_0$;
        \State Empty experience replay buffer;
        \For{$episode=1$ to $E^{\rm out}$}
        \State -------------------- Inner layer Q-learing -------------------- 
        \State Initialize $Q(\mathbf{s}^{\rm inn},\mathbf{a}^{\rm inn})=0$;
        \For{$e = 1$ to $E^{\rm inn}$}
        \State Select an action $\mathbf{a}^{\rm inn}_{e}$ according to the $\epsilon$-greedy policy at state $\mathbf{s}^{\rm inn}_{e}$;
        \If{rand($\cdot$) $< \epsilon$}
        \State Choose the action $\mathbf{a}^{\rm inn}_{e}$ that maximizes the Q-value $Q(\mathbf{s}^{\rm inn}_{e},\mathbf{a}^{\rm inn}_{e})$;
        \Else
        \State Randomly select an action $\mathbf{a}^{\rm inn}_{e}$;
        \EndIf
        \State Execute action $\mathbf{a}^{\rm inn}_{e}$, and calculate the reward value $r(\mathbf{s}^{\rm inn}_{e},\mathbf{a}^{\rm inn}_{e})$;
        \State Update $Q(\mathbf{s}^{\rm inn}_{e},\mathbf{a}^{\rm inn}_{e})$ by (\ref{eq25});
        \EndFor
        \State Obtain the computation offloading decisions $\boldsymbol{\alpha}$ based on the result of the previous episode;
        \State --------------------- Outer layer DDPG ---------------------
        \State Obtain the initial state $\mathbf{s}^{\rm out}_1$;
		\For {training round $t$ = 1 to $T$}
		\State Update the state $\mathbf{s}^{\rm out}_t$;
		\State Select the action $\mathbf{a}^{\rm out}_t=\mu(\mathbf{s}^{\rm out}_t|\theta^\mu)+\beta$ by the online actor network;
		\State Obtain the next state $\mathbf{s}^{\rm out}_{t+1}$ and the reward $r_t$ based on the state $\mathbf{s}^{\rm out}_t$ and action $\mathbf{a}^{\rm out}_t$;
		\State Store the experience $(\mathbf{s}^{\rm out}_t,\mathbf{a}^{\rm out}_t,r_t,\mathbf{s}^{\rm out}_{t+1})$ in $\mathbf{O}$;
        \State Random sample a mini-batch of $o$ experiences $(\mathbf{s}^{\rm out}_j,\mathbf{a}^{\rm out}_j,r_j,\mathbf{s}^{\rm out}_{j+1})$ from $\mathbf{O}$;
        \State Update the critic network with (\ref{eq21}) and (\ref{eq22});
        \State Update the actor network with (\ref{eq18}) and (\ref{eq19});
		\State Update the weights $\theta^{\hat{\mu}}$ and $\theta^{\hat{Q}}$ by (\ref{eq23}) and (\ref{eq24});
		\EndFor
        \EndFor
		\State Return $\mathbf{L}^{\rm usv}, \boldsymbol{\alpha}$.
	\end{algorithmic}  
\end{algorithm} 
\subsection{Computational Complexity Analysis}
In this subsection, we discuss the computational complexity of the proposed DLRL algorithm. For the inner layer, the computational complexity of the Q-learning depends upon the state space size and action space size. Thus, the computational complexity is $\mathcal{O}(E^{\rm inn}|\mathbf{s}^{\rm inn}||\mathbf{a}^{\rm inn}|)$, where $E^{\rm inn}$ is the total number of episodes in the training process, $|\mathbf{s}^{\rm inn}|$ is the state space size, and $|\mathbf{a}^{\rm inn}|$ is the action space size. For the outer layer, the computational complexity of the DDPG usually depends on the actor and critic networks. The actor network has $L_a$ layers with $F_i$ neurons in the $i$-th layer ($i \leq L_a$). The complexity of the $i$-th layer is $\mathcal{O}(F_{i-1}F_i+F_iF_{i+1})$. The complexity of the actor network is $\mathcal{O}(\sum_{i=2}^{L_a-1}(F_{i-1}F_i+F_iF_{i+1}))$ \cite{r64}. The critic network has $L_c$ layers with $G_j$ in the $j$-th layer ($j \leq L_c$). The complexity of the $j$-th layer is $\mathcal{O}(G_{j-1}G_j+G_jG_{j+1})$. The complexity of the critic network is $\mathcal{O}(\sum_{j=2}^{L_c-1}(G_{j-1}G_j+G_jG_{j+1}))$. Thus, the complexity of the outer layer algorithm is denoted as $\mathcal{O}(E^{\rm out}T(\sum_{i=2}^{L_a-1}(F_{i-1}F_i+F_iF_{i+1})+\sum_{j=2}^{L_c-1}(G_{j-1}G_j+G_jG_{j+1})))$. Thus, the overall complexity of the DLRL algorithm is $\mathcal{O}(E^{\rm out}T(\sum_{i=2}^{L_a-1}(F_{i-1}F_i+F_iF_{i+1})+\sum_{j=2}^{L_c-1}(G_{j-1}G_j+G_jG_{j+1}))+E^{\rm out}E^{\rm inn}|\mathbf{s}^{\rm inn}||\mathbf{a}^{\rm inn}|)$.

\section{Performance Evaluation} \label{V}
\begin{table}[t]
\renewcommand{\arraystretch}{1.2}
	\caption{Simulation Parameters}
	\begin{center}
		\begin{tabular}{|p{5.6cm}|p{2.2cm}<{\centering}|}
		\hline
		\textbf{Parameter} & \textbf{Value}\\
        \hline
        Deployment area & 10nmile$\times$ 10nmile \\
        \hline
        The number of vessels, $M$ & 20$\sim$45 \\
        \hline
        The number of USVs, $U$ & 10$\sim$36 \\
        \hline
		The number of tasks per vessel, $I_m$ & 5$\sim$10\\
		\hline
        The computation resource of USV $u$, $C_u$ & 3GHz\\
		\hline
        The computation resource of the cloud center, $C_0$ & 100GHz\\
        \hline
		Channel bandwidth, $B$ & 1MHz\\
		\hline
		Noise power spectral density, $\sigma^2$ & -114dBm\\
		\hline
		The carrier wavelength, $l$ & 0.05m\\
		\hline
        The antenna height of USV $u$, $H_u$ & 5m \\
        \hline
		The antenna height of vessel $m$, $H_m$ & 9.8m \\  
        \hline
        The transmit power of USV $u$, $P_u$ & 44.7dBm \\
		\hline
		The transmit power of vessel $m$, $P_m$ & 47dBm \\
        \hline
		\end{tabular}
		\label{tab2}
	\end{center}
\end{table}

We develop a simulator in Python to evaluate the performance of the proposed DLRL framework for an MEC-assisted maritime wireless network with multiple USVs as described in Section~\ref{III}. The DLRL algorithm is implemented in TensorFlow. we set $\gamma_{\mu}=0.0001$ and $\gamma_{Q}=0.0001$ as the learning rate for the actor network and the critic network, respectively, and $\gamma^{\rm inn} = 0.01$ as the learning rate for the inner layer. Both the actor network and the critic network have three hidden layers with $(256,128,64)$ neurons. The main simulation parameters used in the experiments are listed in Table~\ref{tab2}, as in~\cite{r1,r39,r51}.

We also implement four benchmark algorithms for performance comparison as follows.

\begin{enumerate}
    \item Particle Swarm Optimization (PSO) and Greedy (PSO-G) algorithm~\cite{r78}. We select PSO algorithm as the first benchmark algorithm for performance comparison as it provides a generic solution to handle complex optimization problems with mixed variables. Specifically, a PSO algorithm based on swarm intelligence search is adopted to optimize the locations of USV clusters; and a greedy algorithm is used to optimize the computation offloading decisions. The USV cluster deployment and computation offloading decisions are updated iteratively until a convergence is reached.

    \item PSO and Double Deep Q-Network (PSO-DDQN) algorithm~\cite{r92}. We selected the PSO-DDQN algorithm as the second benchmark due to its effectiveness in addressing MINLP problems. Specifically, the PSO algorithm is employed to optimize the locations of USV clusters, while the DDQN algorithm is utilized to refine the computation offloading decisions. This combined approach optimizes the computation offloading strategies based on the USVs' locations, iteratively updating until convergence is achieved.

    \item Deep Q Network (DQN) algorithm~\cite{r77}. A deep reinforcement learning algorithm based on Deep Q-Network (DQN) is selected as the third benchmark algorithm due to its high performance in tackling problems with discrete action space. In order to handle mixed action spaces, the actions of USV deployment are discretized, while the action of offloading is set as that in Section~\ref{IV-A}.
    
    \item DDPG algorithm. A deep reinforcement learning algorithm based on DDPG is selected as the fourth benchmark algorithm due to its high performance in tackling problems with continuous action space~\cite{r85}. To handle the mixed action spaces, the discrete offloading action is approximated in continuous action space, while the continuous actions of USV deployment are set as that in Section~\ref{IV-A}.
\end{enumerate}

\begin{figure}[t]
  \centering
  \includegraphics[width=0.5\textwidth]{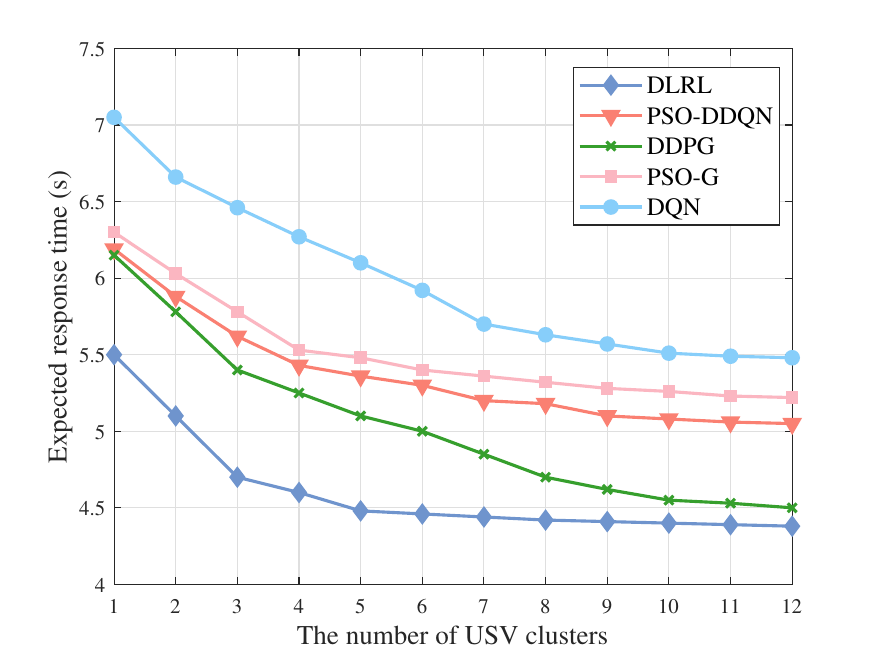}\\
  \caption{Performance comparison of the expected response time ($M=40$ vessels and $U=24$ USVs).}\label{clu}
\end{figure}

Fig.~\ref{clu} compares the performance of the proposed algorithm with that of the four benchmark algorithms versus different numbers of USV clusters. The number of vessels and USVs are set to $M=40$ and $U=24$. The computation resource of each USV is $3$~GHz. Each vessel carries out a variable number of computation tasks from $5$ to $10$. It is shown that the proposed learning algorithm outperforms the PSO-G, PSO-DDQN, DQN, and DDPG in terms of the achieved response time. The DDPG uses continuous actions to approximate the discrete offloading decisions, while the DQN algorithm discretizes the actions of USV deployment, resulting in reduced accuracy, compared with the proposed dual-layer learning solution. The PSO-G and PSO-DDQN algorithms involve iterative optimization for both discrete and continuous variables. Generally, PSO-G prioritizes exploration to find the optimal solutions while DDPG and DQN algorithms prioritize exploitation by leveraging learned policies to maximize the rewards. PSO-DDQN algorithm as a form of deep reinforcement learning by incorporating PSO and DDQN for exploration and exploitation. The proposed DLRL algorithm that integrates DDPG and Q-learning to handle the continuous and discrete actions achieves the best performance. It can also be seen that the expected response time decreases when the number of USV clusters increases from $1$ to $5$, and the decrease rate becomes much slower for $N>5$. As the number of USV clusters increases, the distance between the vessel and the USV cluster decreases, which effectively decreases the transmission time. The results suggest that it may not be necessary to deploy each individual USV to the optimal location, and an appropriate number of USV clusters can be determined to facilitate the efficient deployment without notably reducing the network performance. 

\begin{figure}[t]
    \centering
    \subfigure[Edge computing ratio]{
        \includegraphics[width=0.44\textwidth]{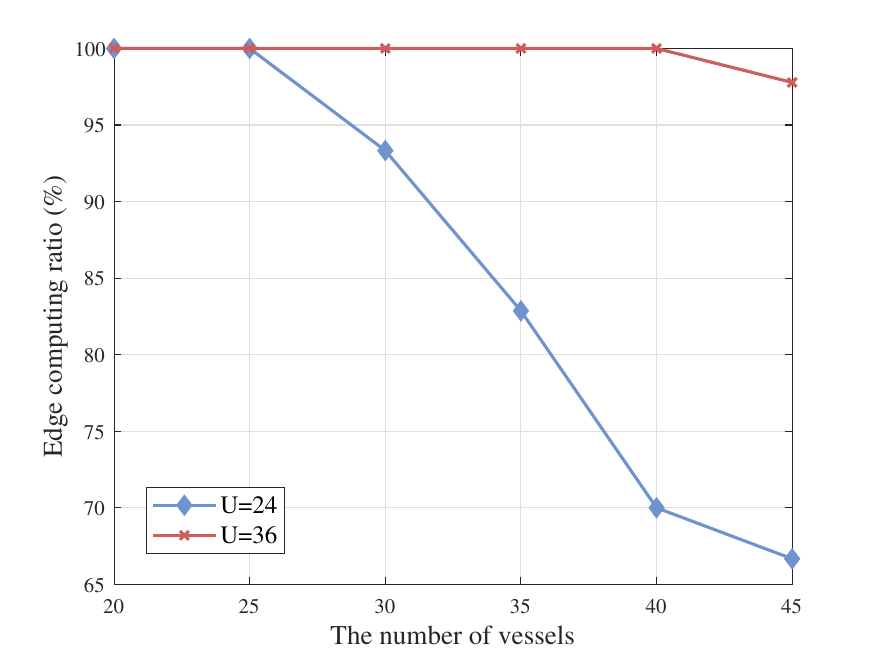}
    }
    \subfigure[Expected response time ($U=24$ USVs)]{
        \includegraphics[width=0.44\textwidth]{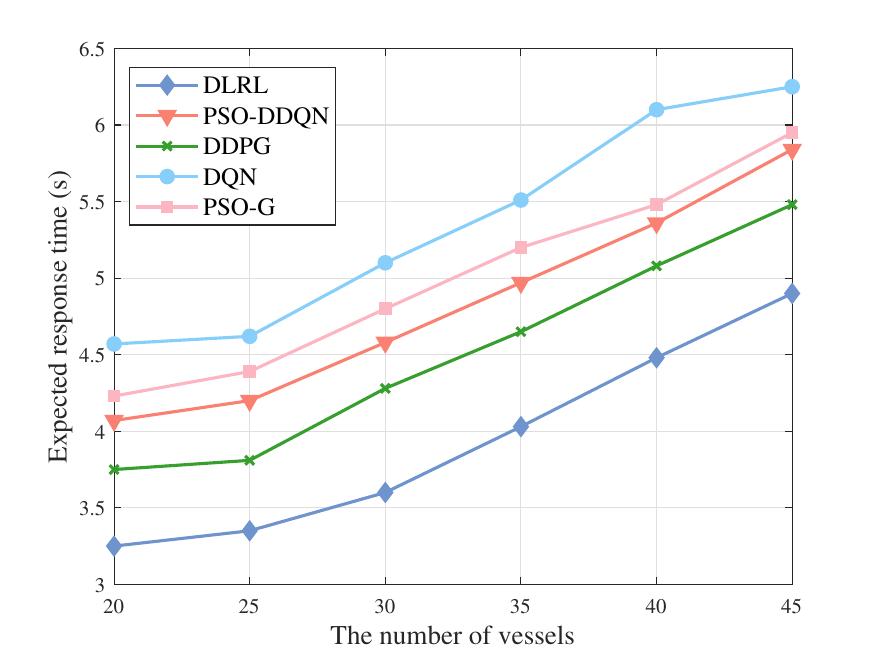}
    }
    \subfigure[Expected response time ($U=36$ USVs)]{
        \includegraphics[width=0.44\textwidth]{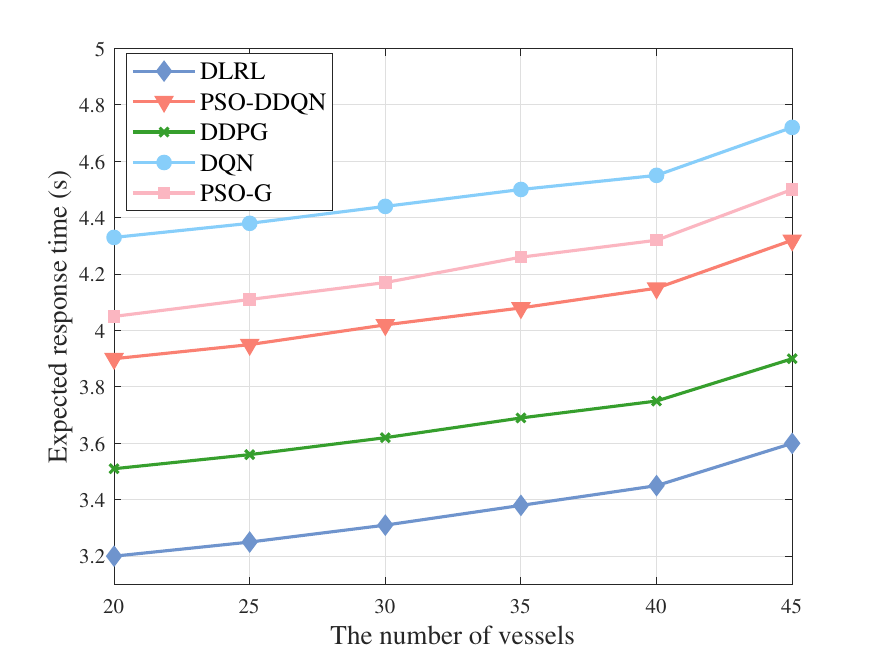}
    }
    \caption{Performance study of the edge computing ratio and expected response time ($N=5$ USV clusters).}
    \label{fig6}
\end{figure}

Fig.~\ref{fig6} studies the performance of the proposed algorithm under different numbers of tasks. The edge computing ratio, defined as the number of computation tasks offloaded to the USVs for edge computing over the total number of tasks, is shown in Fig.~\ref{fig6} (a). For $U=24$ and $N=5$, when the number of vessels exceeds $M>25$, some tasks cannot be offloaded to the edge server but have to be forwarded to the remote cloud for data processing, due to the limited computation resources of deployed USVs. Thus, the edge computing ratio decreases with the number of vessels, and the corresponding expected response time increases as shown in Fig.~\ref{fig6} (b). We further increase the number of USVs to $U=36$, and more computation tasks can be offloaded to the USVs with more computation resources, and the edge computing ratio is higher than that with $U=24$, and a smaller expected response time is achieved correspondingly, as shown in Fig.~\ref{fig6} (c). 

\begin{figure}[t]
    \centering
    \subfigure[Edge computing ratio]{
        \includegraphics[width=0.5\textwidth]{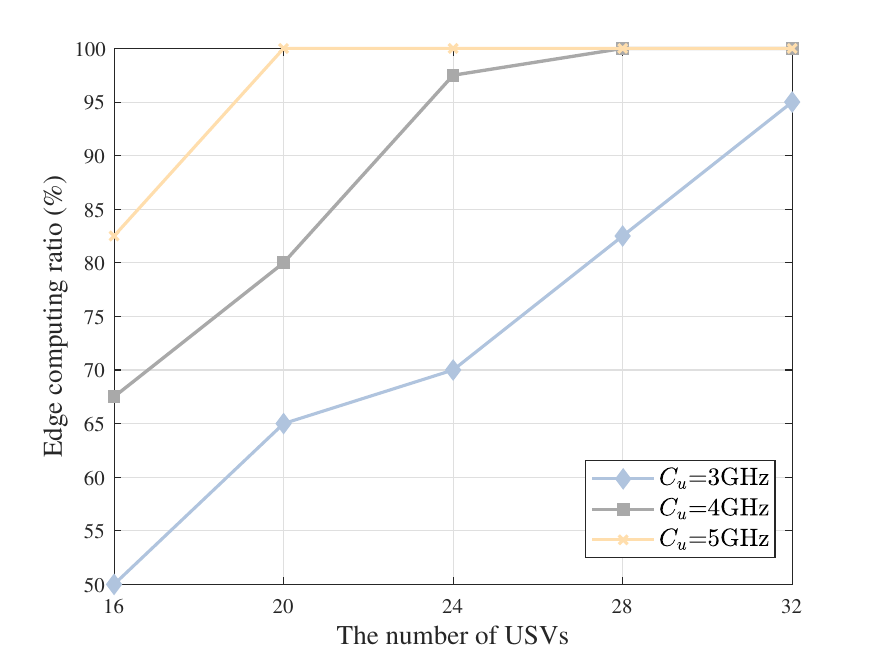}
    }
    \subfigure[Expected response time]{
        \includegraphics[width=0.5\textwidth]{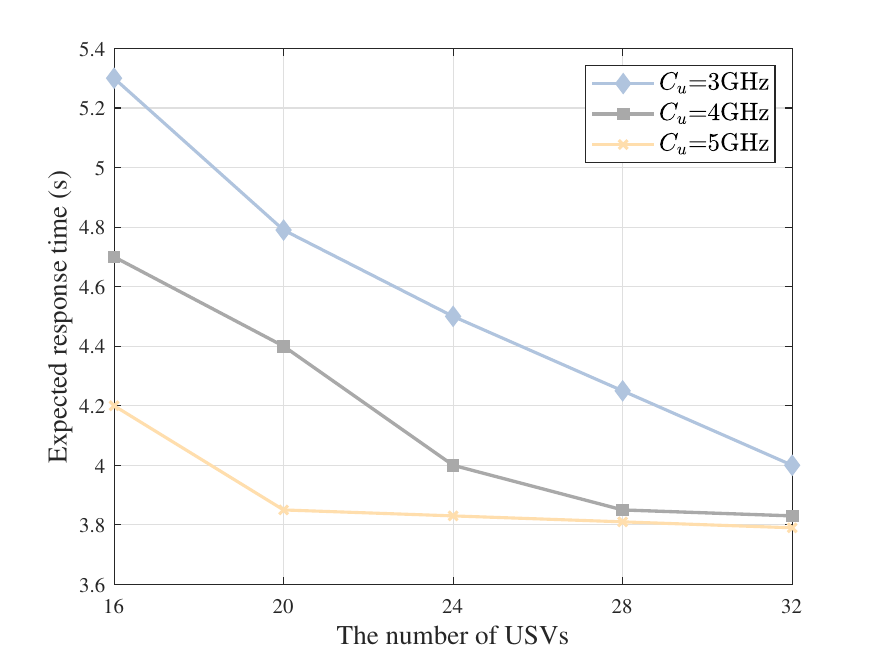}
    }
    \caption{The impact of the computation resources ($M=40$ vessels and $N=5$ USV clusters).}
    \label{fig8}
\end{figure}

Fig.~\ref{fig8} studies the impact of the computation resources on the performance of the proposed algorithm. The more computation resources, the more tasks can be offloaded to the edge server to achieve a lower response time. As shown in Fig.~\ref{fig8}, when the computation resource is larger than $100,$ i.e., in the case of $U>20$ and $C_u=5{\rm GHz}$, or $U>28$ and $C_u=4{\rm GHz},$, all computation tasks of vessels can be offloaded to the edge servers to attain the minimum expected response time. 

\begin{figure}[t]
    \centering
    \subfigure[$M=15$ vessels, $U=10$ USVs]{
        \centering
        \includegraphics[width=0.5\textwidth]{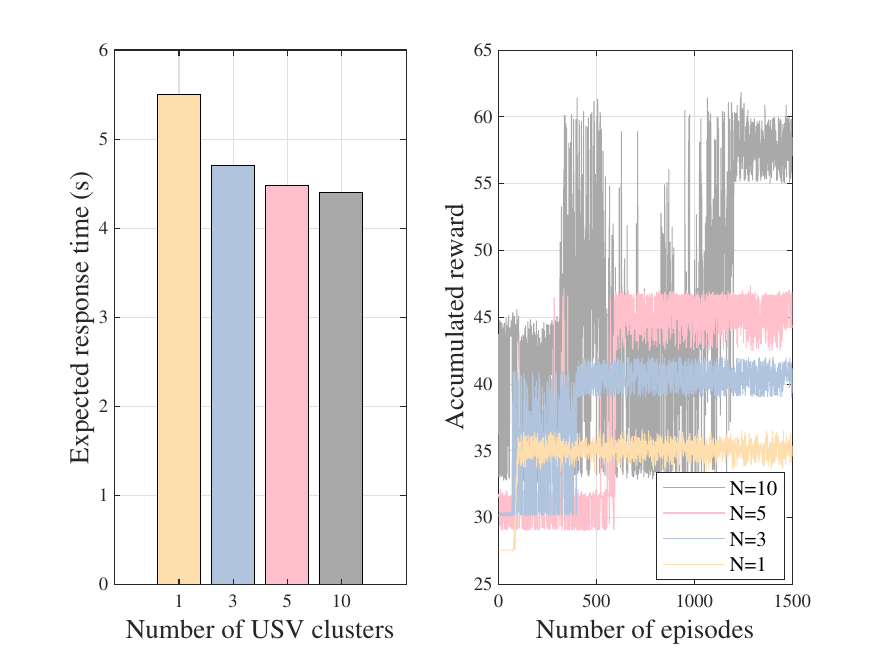}
    }
    \subfigure[$M=24$ vessels, $U=40$ USVs]{
        \centering
        \includegraphics[width=0.5\textwidth]{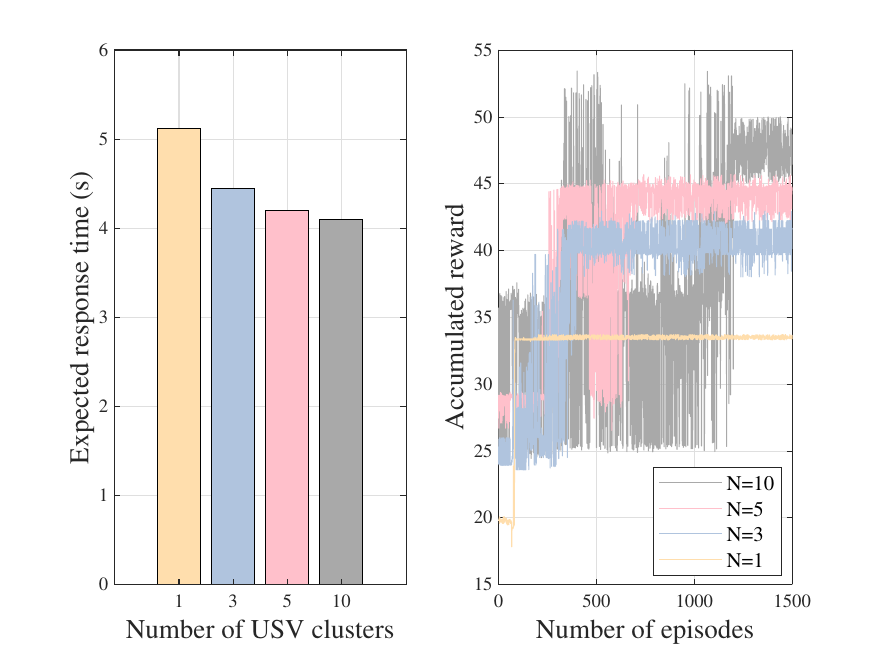}
    }
    \caption{Convergence performance versus different numbers of USV clusters and vessels.}\label{fig4}
\end{figure}

The convergence performance of the proposed DLRL algorithm under different numbers of USV clusters and vessels is shown in Fig.~\ref{fig4}. In Fig.~\ref{fig4} (a), the number of vessels and USVs are set to $M=15$ and $U=10$, respectively. It is found that the achieved expected response time for $N=10$ is reduced compared with the cases of $N=1,3,5$ clusters. It is also observed that the convergence time increases with $N$ due to a larger searching space of the optimal locations. For example, when the number of USV clusters is $N=10$, the convergence performance increases by about $1100$, $800$, $600$ episodes compared with the cases of $N=1,3,5$, respectively. Similar results can be observed in Fig.~\ref{fig4} (b) where the number of vessels and USVs increases to $M=40$ and $U=24$, respectively. Compared with the case of $N=5$, the expected response time for $N=10$ is reduced by $1.8\%$ at the cost of increased convergence time. In a large scale IoV networks with a high number of vessels and USVs, clustering provides an efficient and scalable solution to strike a balance between the response time and the convergence time.

\begin{figure}
  \begin{center}
  \includegraphics[width=0.5\textwidth]{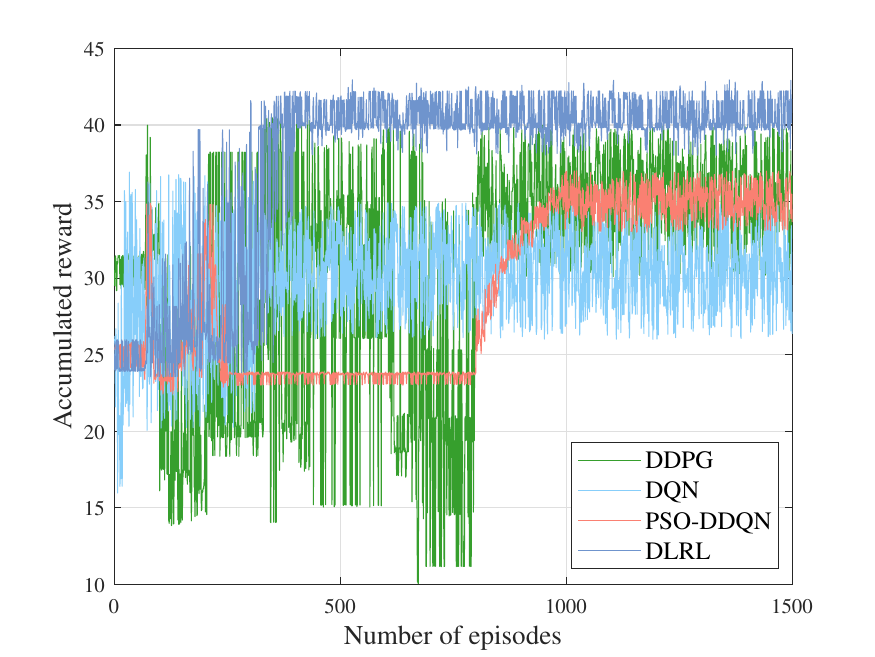}\\
  \caption{Convergence performance of DLRL, PSO-DDQN, DDPG and DQN algorithms.}\label{fig3}
  \end{center}
\end{figure}

Fig.~\ref{fig3} compares the convergence performance of the proposed four learning algorithms, i.e., DLRL, PSO-DDQN, DDPG, and DQN~\footnote{The converge performance of PSO-G is not presented as PSO-G is not typically based on rewards as the reinforcement learning algorithms.}. The proposed DLRL algorithm typically converges in about $400$ episodes, while the DDPG algorithm requires approximately $800$ episodes to converge. In addition, the final accumulated reward of the DLRL algorithm is slightly higher than that of the DDPG algorithm because the DDPG approximates the discrete offloading action in continuous action space, resulting in reduced accuracy. Although the DQN algorithm achieves a good convergence performance, the accumulated reward learned from the DQN is lower than that of the DLRL, as DQN discretizes the continuous action space which results in performance degradation. For the PSO-DDQN algorithm, the convergence performance is inferior to that of the DLRL, as the incorporation of PSO increases the complexity of exploration, which ultimately degrades overall performance. The proposed DLRL algorithm achieves the highest rewards with good convergence performance compared to PSO-DDQN, DDPG and DQN primarily due to its effective handling of hybrid actions.

\section{Conclusion} \label{VI}
In this paper, we have studied the mobile edge deployment and resource management for MEC-assisted maritime wireless networks, where USVs with computation resources are deployed to provide edge computation services for vessels. A dual-layer reinforcement learning framework has been proposed to optimize the deployment of USVs and computation offloading decisions. Simulation results have demonstrated that the USV cluster deployment provides an efficient and scalable solution for large scale maritime wireless networks with a high number of USVs and vessels. Compared with four benchmark algorithms in the literature, the proposed DLRL algorithm has achieved a significant improvement in reducing the expected response time. In the future, we will further study the computation offloading in a heterogeneous MEC-assisted maritime wireless network with different types of tasks with different QoS requirements. 


\ifCLASSOPTIONcaptionsoff
  \newpage
\fi

\bibliographystyle{IEEEtran}
\bibliography{IEEEabrv,Bibliography}

\vfill

\end{document}